\documentclass[doublecol]{epl2} 

\usepackage{amsmath}
\usepackage{amssymb,amsthm}

\renewcommand{\vec}[1]{\mathbf{#1}}

\newcommand{\ii}{\mathrm{i}}

\newcommand{\mean}[1]{\langle{#1}\rangle}

\newcommand{\ket}[1]{|{#1}\rangle}

\newcommand{\fig}[1]{fig.~\ref{#1}}

\title{Field-assisted doublon manipulation in the Hubbard model. A quantum doublon ratchet}
\shorttitle{Field-Assisted Doublon Manipulation in the Hubbard Model}

\author{K. Balzer \and M. Eckstein}
\shortauthor{K. Balzer \etal}

\institute{                    
  Max Planck Research Department for Structural Dynamics - University of Hamburg-CFEL, Building 99, Luruper Chaussee 149, 22761 Hamburg, Germany
  }


\abstract{
For the fermionic Hubbard model at strong coupling, we demonstrate that directional transport of 
localized doublons (repulsively bound pairs of two particles occupying the same site of the crystal 
lattice) can be achieved by applying an unbiased ac field of time-asymmetric (sawtooth-like) shape. 
The mechanism involves a transition to intermediate states of virtually zero double occupation which 
are reached by splitting the doublon by fields of the order of the Hubbard interaction. The process
is discussed on the basis of numerically exact calculations for small clusters, and we apply it to more 
complex states to manipulate the charge order pattern of one-dimensional systems.
}

\begin{document}

\maketitle

To control and tune macroscopic properties by directly manipulating processes on the 
atomic scale is an ultimate goal in condensed matter physics. For example, one can
selectively excite definite phonon modes~\cite{kaiser:13} and thus create 
long-lived transients which exhibit interesting properties like superconductivity~\cite{fausti:11}. 
The design of suitable mechanisms that support such objectives requires to understand
the underlying interplay of strong fields and many-particle interactions, which, nowadays, 
can be simulated and studied for more and more diverse situations by experiments with 
ultracold atoms~\cite{jordens:08,bloch:12,windpassinger:13}. The unprecedented control 
over the parameters in those systems allows one to explore the crossover between few 
and many-particle physics~\cite{wenz:13} and to probe or manipulate systems with 
single-site resolution~\cite{endres:13}. This makes cold atoms well suited to analyze 
field-induced processes all the way from addressing single particles to complex many-particle 
states. In extended systems, the impact of an electric field on the correlated particle
motion is already nontrivial at weak to moderate interactions (leading, {\em e.g.}, to damping 
of Bloch oscillations~\cite{freericks:08,eckstein:11.bloch.osc} or nonlinear transport~\cite{okamoto:08}). 
However, it becomes even more subtle if both the interactions and the field are comparable to or larger than the bandwidth. 
In this regime, fields can lead to the dielectric breakdown of a Mott 
insulator~\cite{oka:10,eckstein:10.dielectric.breakdown,lenarcic:12} (which is the many-body
analog of the Zener breakdown in band insulators), and when the field and the interaction are resonant, the coupling of many degenerate states 
can lead to the emergence of phases which are described in terms of effective spin
models~\cite{sachdev:02,simon:11,meinert:13}.

In this Letter, we investigate a controlled manipulation of few-particle states based on 
the interplay of the local repulsive interaction and the external field. 
We focus on the single-band Fermi-Hubbard model, which is the paradigm model for strongly 
correlated systems such as cold atoms in optical lattices or interacting electrons in a solid, and we 
develop a protocol by which an asymmetric alternating (ac) field is used to translate a repulsively bound 
``doublon''~\cite{winkler:06} along a chain in a directional motion, reminiscent of a (nondissipative) 
ratchet~\cite{denisov:13}. In the case of a {\em single particle} on a tight binding chain, an unbiased 
time-periodic driving cannot lead to a finite current or set an initially localized particle into motion, 
not even when the driving is asymmetric in time and thus breaks time-reversal symmetry~\cite{goychuk:01,ponomarev:09}. 
By contrast, the rectification effect in our protocol relies on the {\em interaction} between the two particles 
that form the doublon, which will also allow us to use the protocol for a controlled transmutation of more 
complex states. Moreover, the mechanism can be viewed as a generalization of an optimal control problem 
which has been discussed for a double quantum dot system~\cite{blasi:13} and is in line 
with charge transfer effects in molecules which are exposed to asymmetric laser fields~\cite{tagliamonti:13}.

The Hamiltonian is given by
\begin{equation}
\label{eq:ham}
H(t)=H_J+H_U+H_F(t)\;,
\end{equation}
where \mbox{$H_J=-J\sum_{\langle ij\rangle\sigma} c_{i\sigma}^\dagger c_{j\sigma}$} describes the 
hopping of fermions with spin $\sigma$ between nearest-neighbor sites $i$ and $j$, 
\mbox{$H_U=U\sum_{i} n_{i\uparrow} n_{i\downarrow}$} accounts for the local 
repulsive interaction, and \mbox{$H_F(t)=\sum_{i\sigma} [\vec{F}(t)\cdot\vec{r}_i] n_{i\sigma}$} is the coupling to an
external field $\vec{F}(t)$ ($n_{i\sigma}=c^\dagger_{i\sigma}c_{i\sigma}$ is the density, and the lattice spacing is chosen as unit of length, such that fields are measured in units of energy).
In the Hubbard model, two particles with opposite spin on the same site form a doublon for 
sufficiently strong interaction $U$: For $U>2W$, where $W$ is the single-particle bandwidth, 
a single doublon in an otherwise empty lattice can no longer decay into two particles and thus 
forms a repulsively bound pair~\cite{winkler:06}, which has an infinite lifetime (at zero external field) 
and an effective hopping amplitude $\propto J^2/U$~\cite{hofmann:12}. 

\begin{figure}
\includegraphics[width=0.485\textwidth]{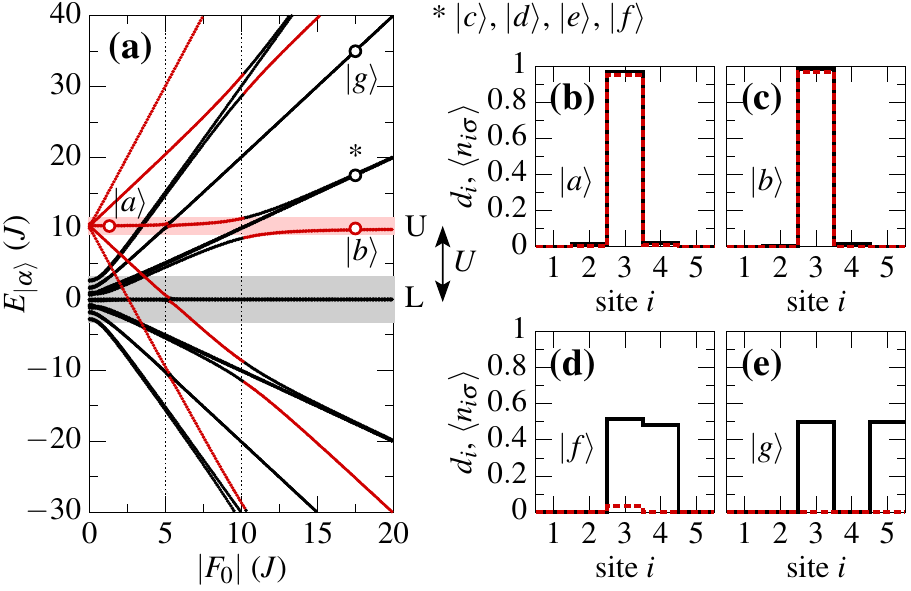}
\caption{(a)~Stark-shifted energy levels $E_{\ket{\alpha}}$ for two fermions of 
opposite spin on a five-site chain as function of $|F_0|$ at $U=10J$. The labels L and U refer 
to the lower and upper Hubbard bands which strongly overlap at large fields. States with double 
occupation $d>0.5$ [$d=\sum_id_i=\sum_i\mean{n_{i\uparrow}n_{i\downarrow}}$] are plotted in red. 
Panels~(b)-(e) show the densities $\mean{n_{i\sigma}}$ (black solid lines) and 
the local double occupations $d_i$ (red dashed lines) of selected states $\ket{\alpha}$ in~(a).}
\label{fig:fig1}
\end{figure}

The manipulation of doublons in the field can be understood by analyzing the Stark effect for two particles 
in a Hubbard chain. Figure~\ref{fig:fig1}a shows the Stark-shifted many-particle energy levels 
for such a system at $U=10J$, obtained by diagonalizing $H$ for $\vec{F}(t)=F_0\vec{e}_x$.
At zero field, the strong on-site interaction splits the spectrum into two bands, well separated by 
an energy gap $U$. The lower band corresponds to the ``continuum'' of states with two separated 
particles (double occupancy $d\ll1$), while the upper band belongs to the repulsively bound doublon 
(double occupancy $d \approx 1$). In a finite field $F_0$, both bands further split 
into a Wannier-Stark ladder~\cite{wannier:60} with states spaced by approximately $F_0$ (for the lower band) or 
$2F_0$ (for the upper band). For sufficiently strong fields, the Stark effect thus leads to essential mixing of
states of high and low double occupation. For $J=0$, the crossing occurs at characteristic 
field strengths $|F_0|=U/n$ with integer $n\geq1$, when the doublon on site $i$ is resonant 
with two separated particles on sites $i$ and $i+n$ (for $F_0>0$) or on sites $i$ and $i-n$ (for $F_0<0$). 
For $J\neq0$, these resonant levels hybridize and turn into avoided crossings.

In the following, we consider the states labeled~$\ket{a}$ to $\ket{g}$ in \fig{fig:fig1}a 
($F_0$ finite). Both states~$\ket{a}$ and~$\ket{b}$ describe a doublon $\ket{D_3}=\ket{\uparrow_3\downarrow_3}$ 
which is localized by the field on the central site $3$ of the chain, see figs.~\ref{fig:fig1}b 
and~\ref{fig:fig1}c. The states $\ket{c}$ to $\ket{f}$ are very close in energy, become fully 
degenerate for $|F_0|\rightarrow\infty$ and have low double occupancy. The energetically lowest and highest 
states in this manifold correspond to the antisymmetric and symmetric superpositions 
$\frac{1}{\sqrt{2}}(\ket{\uparrow_3\downarrow_4}-\ket{\downarrow_3\uparrow_4})$ 
and $\ket{S_{34}}=\frac{1}{\sqrt{2}}(\ket{\uparrow_3\downarrow_4}+\ket{\downarrow_3\uparrow_4})$,
respectively [see $\ket{f}$ in \fig{fig:fig1}d]. In between we find states of the form 
$\frac{1}{\sqrt{2}}(\ket{\uparrow_2\downarrow_5}\pm\ket{\downarrow_2\uparrow_5})$
which pass straight through the avoided level crossing at $|F_0|=10J$. Finally, $\ket{g}$ is
the superposition $\ket{S_{35}}=\frac{1}{\sqrt{2}}(\ket{\uparrow_3\downarrow_5}+\ket{\downarrow_3\uparrow_5})$, 
see \fig{fig:fig1}e, which, like $\ket{f}$, has practically zero double occupancy. Now suppose 
we prepare the system in the doublon state $\ket{a}$ at a weak field $F_0$. If the field is then ramped up 
to $|F_0|\gg U$ on a fast time scale, the local doublon is preserved, and the system is diabatically transferred 
into state $\ket{b}$. Contrarily, if the field is turned on smoothly, different final states can be 
reached depending on which avoided crossing is followed adiabatically. For example, if we quickly pass the 
crossings at $|F_0|=U/n$ for $n>1$ and then slowly switch through the resonance at $|F_0|=10J$ we 
will end up in state $\ket{S_{34}}$. Similarly we can transfer the system into state $\ket{S_{35}}$
by switching adiabatically around the $n=2$ crossing at $|F_0|=5J$ (note that the much narrower 
crossing requires an essentially slower field tuning in this case).

\begin{figure}
\includegraphics[width=0.485\textwidth]{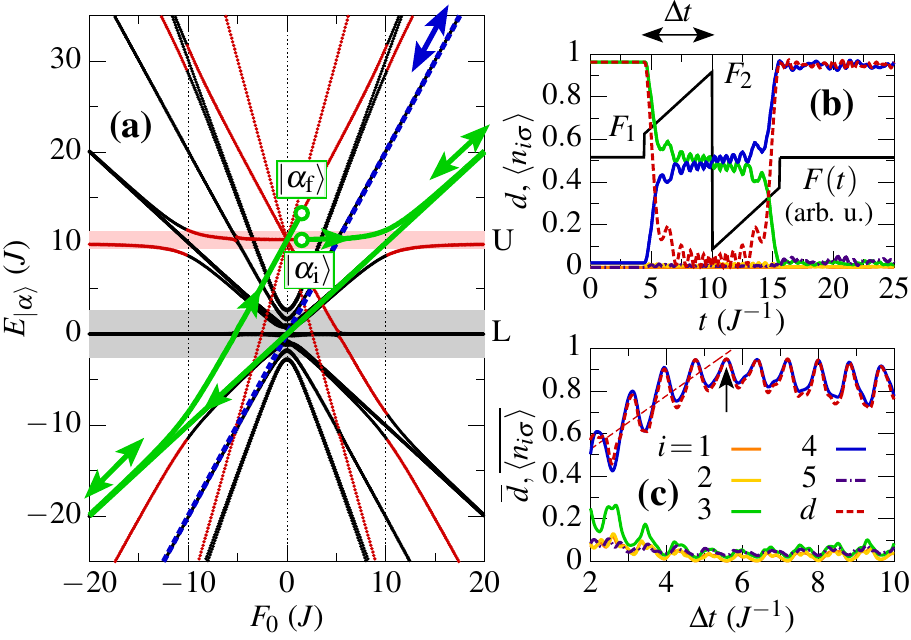}
\caption{Doublon manipulation on the linear chain of \fig{fig:fig1}; $U=10J$. 
(a)~Illustration of the applied field cycle (green arrows). (b)~Unitary time evolution of $d(t)$ 
and $\mean{n_{i\sigma}}(t)$ for a doublon initially (at time $t=0$) prepared on the central site 
with $F(0)=F_0=J$. The black line shows the field $F(t)$ shifted in vertical direction and arbitrarily rescaled. In eq.~(\ref{eq:field}), we 
set $F_1=7.5J$, $F_2=25J$, $t_0=10J^{-1}$ and $\Delta t=5.584J^{-1}$ (cf.~arrow in panel (c)). 
(c)~Dependence of the final values of $d$ and $\mean{n_{i\sigma}}$ on the ramp time $\Delta t$ 
(values are time-averaged over a period of $10J^{-1}$).}
\label{fig:fig2}
\end{figure}

Including positive and negative fields, a repeated swap between states of the form $\ket{D_i}$ 
and $\ket{S_{ij}}$ can now be exploited to manipulate doublons in a time-dependent fashion: The 
field cycle indicated by the green arrows in \fig{fig:fig2}a, {\em e.g.}, moves a doublon 
on site $3$ (label $\ket{\alpha_\mathrm{i}})$ by one lattice spacing to the right (label $\ket{\alpha_\mathrm{f}}$). 
As a definite realization of the cycle we choose a time-dependent field $\vec{F}(t)=F(t)\vec{e}_x$ of the form
\begin{equation}
\label{eq:field}
F(t)=
\left\{
\begin{array}{ll}
F_1+\gamma (t-t_0^-)    &, t_0^-< t\leq t_0\\
-F_2+\gamma (t-t_0)   &, t_0< t\leq t_0^+\\
F_0                   &,\textup{otherwise}
\end{array}
\right.\;,
\end{equation}
with $\gamma=(F_2-F_1)/\Delta t$ and $t_0^\pm=t_0\pm\Delta t$ (fig.~\ref{fig:fig2}b).
First, irrelevant level crossings are passed by the sudden change of the electric field 
from $F_0$ to $F_1$, and the field is ramped linearly (within a time interval $\Delta t$) from 
$F_1$ to $F_2$  through the $n=1$ resonance to split the doublon. Thereafter the field is suddenly
reversed (leaving the state invariant), and the process is traversed in opposite manner to 
restore the doublon on site $4$.

\begin{figure}
\includegraphics[width=0.485\textwidth]{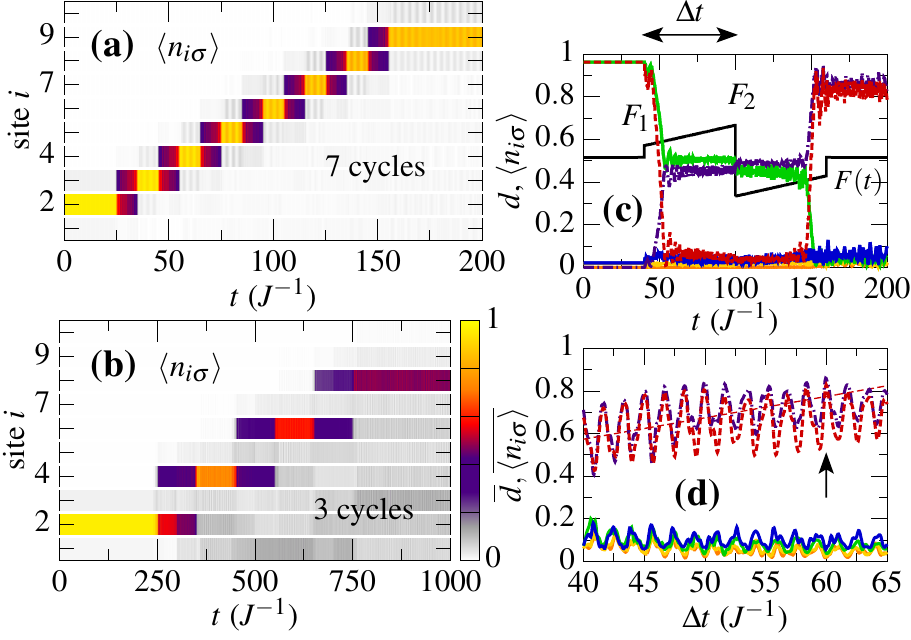}
\caption{(a) and~(b): Field-assisted directional transport of an initially localized doublon 
through a one-dimensional chain with ten sites at $U=10J$ by one site per field cycle (panel~(a)) and two sites per field cycle (panel~(b)). 
After seven (three) cycles, the total double occupation is decreased from $0.96$ to $0.87$ ($0.62$) in panel~(a) (panel~(b)); 
the field parameters are as in \fig{fig:fig2}b (\fig{fig:fig3}c). (c) and~(d): Same as in figs.~\ref{fig:fig2}b and \ref{fig:fig2}c 
but for the translation of the doublon by two sites at a time; parameters: $F_0=J$, $F_1=4.25J$, $F_2=10J$ 
and, in panel (a), $\Delta t=59.986J^{-1}$.}
\label{fig:fig3}
\end{figure}

In \fig{fig:fig2}b, we numerically validate the field-assisted transport mechanism, 
time-evolving a doublon which is initially located on the central site of a five-site chain~\cite{fn:2}. When 
the field passes the resonance at $F\equiv U=10J$ (during the positive part of the field cycle), 
the double occupancy rapidly drops to $d\approx0$, and the density becomes equally distributed over sites~$3$ 
and~$4$. Within the negative part of the cycle, the initial double occupancy is recovered, and the doublon emerges shifted on site~$4$. 
Figure~\ref{fig:fig2}c shows that the resonance must be passed sufficiently slowly 
($\Delta t>5J^{-1}$) to generate the displaced doublon with high fidelity. The oscillations
in the fidelity can be explained by Bloch oscillations of the intermediate unpaired state 
$\ket{S_{34}}$ at frequency $\omega_\mathrm{B}(t)=|F(t)|$, which are also visible 
in \fig{fig:fig2}b.
Finally, \fig{fig:fig3}a proves that it is possible, by applying multiple cycles of the form 
(\ref{eq:field}), to successively move the doublon site by site through the whole lattice without 
destroying the spatio-temporal coherence.
Note that $F(t)$ has zero time-average, such that any motion of the doublon 
corresponds to a ratchet-like rectification of the field. Changing the sign 
of the field cycle will reverse the transport direction. 

The feasibility to shift a doublon by two sites in a single field cycle, following the blue path 
in \fig{fig:fig2}a via the intermediate state $\ket{S_{35}}$ [state $\ket{g}$ in \fig{fig:fig1}e],
is shown in figs.~\ref{fig:fig3}b-d. As aforementioned, 
a slower ramp is required to remain adiabatic around the narrower crossing at $F=5J$ ($\Delta t$ is about ten times larger), and as 
a result the intermediate state is subject to stronger decay. However, one can still achieve a final double occupancy
of $d\approx0.8$ with optimal field parameters (red dashed line). With the optimal ramp {\em successive} 
motion in steps of two can be realized at least over a few steps although it is much harder than the one-step process, as 
seen from \fig{fig:fig3}b.

\begin{figure}[t]
\includegraphics[width=0.485\textwidth]{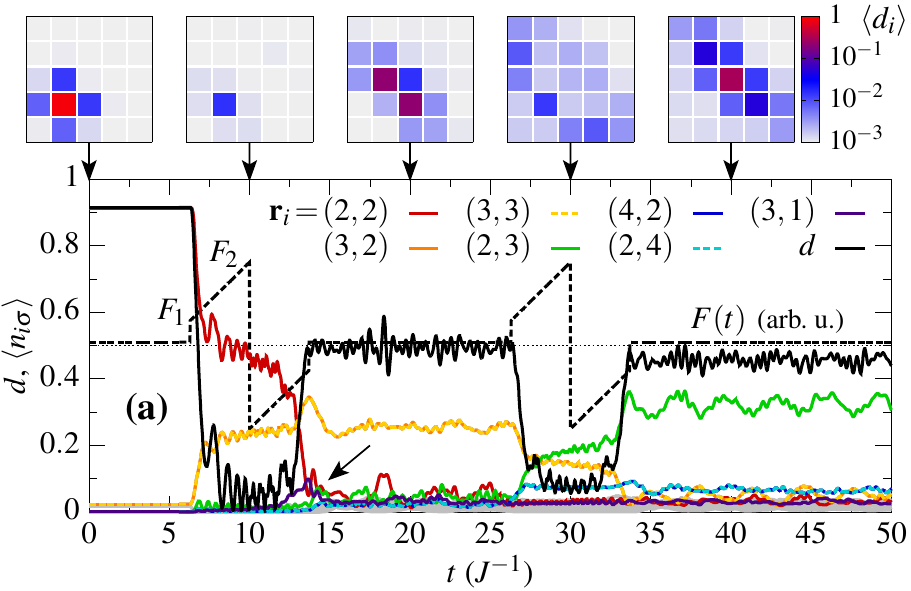}\\
\includegraphics[width=0.485\textwidth]{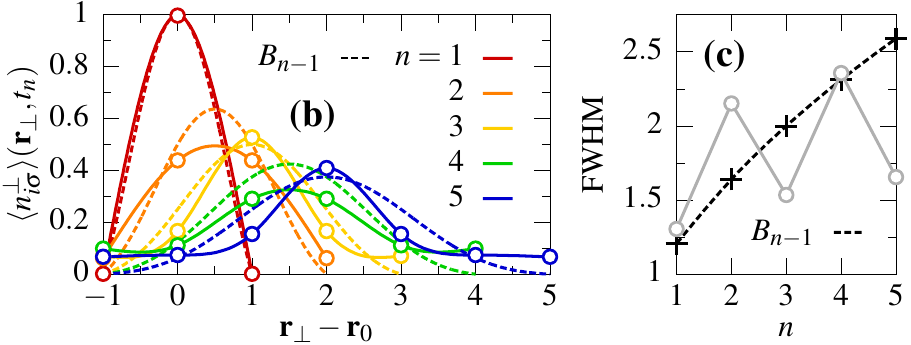}
\caption{(a)~Doublon transport mechanism applied to a square ($5\times5$) lattice with a 
doublon initially localized on the site $\vec{r}=(2,2)$; the field $\vec{F}(t)=F(t)(\vec{e}_x+\vec{e}_y)$ 
[black dashed line] is pointing along the lattice diagonal and $U=10J$. With exception of $\Delta t=3.7J^{-1}$ 
the field parameters are as in \fig{fig:fig2}b. While the colored curves show the time evolution 
of the particle density $\mean{n_{i\sigma}}(t)$ on selected sites $\vec{r}_i$, the black solid line refers to the total double occupancy $d(t)$. 
The panels above the figure show the spatial distribution of the double occupation at 
times $t/J^{-1}=0$, $10$, $20$, $30$ and $40$ (plotted on logarithmic scale).  
(b)~Transverse charge distribution $\mean{n_{i\sigma}^\perp}(\vec{r}_{\perp},t_n)$ after cycle $n$ 
(solid lines) for a calculation, where a single doublon has 
initially been prepared on the site $\vec{r}_0=(2,2)$ of a $7\times7$ cluster. The dashed lines correspond 
to a binomial distribution $B_{n-1}(\vec{r}_\perp)$. (c)~Comparison of 
the full widths at half maximum (FWHM) of the distributions in panel~(b).
}
\label{fig:fig4}
\end{figure}

An interesting question is whether the doublon manipulation protocol can be generalized to
higher dimensions. If the field is applied along a principal axis of the lattice, the 
protocol will not work because particles can delocalize in directions perpendicular
to the field. Although the doublon moves only with a reduced hopping $J^2/U$, 
the intermediate state of two separated particles can delocalize (independent of $U$) on a 
time scale $1/J$ which is not small compared to the traverse of the level crossing. On the other hand, 
if the field is applied along the lattice diagonal, the transverse doublon and single-particle decay 
are strongly suppressed because to displace a particle perpendicular to $\vec{F}_0$ one requires two 
hopping processes against field gradients of $\pm |\vec{F}_0|$, {\em i.e.}, via an off-resonant intermediate 
state. Figure~\ref{fig:fig4} shows the time evolution during two successive field cycles
for a doublon which is initially prepared on the site $\vec{r}=(2,2)$ of a $5\times5$ square
lattice (the parameters $F_{0,1,2}$ are as in \fig{fig:fig2}b and $\Delta t=3.7J^{-1}$). 
Obviously, the first cycle, which is centered at $t=10J^{-1}$, splits the doublon into two equal parts which become 
located on the next-nearest-neighbor sites $(2,3)$ and $(3,2)$. According to this observation, we might 
expect that additional field cycles lead to further transverse fragmentation, such that the doublon 
dynamics resembles the classical diffusion dynamics on a Galton board (quincunx)~\cite{galton:1889} 
which results in a binomial distribution $B_n(\vec{r}_\perp)$ with $n\geq0$. However, 
\fig{fig:fig4} reveals that this picture is not applicable, and quantum interference 
effects are important: After the second cycle (for $t\gtrsim35J^{-1}$) we recover a single doublon 
which is well localized on site $(3,3)$, cf.~the green line. Furthermore, close to $50\%$ of 
the doublon are lost during the first cycle, while the second cycle almost conserves the total doublon weight.

\begin{figure}
{
\hspace*{1pc}\includegraphics[width=0.11\textwidth]{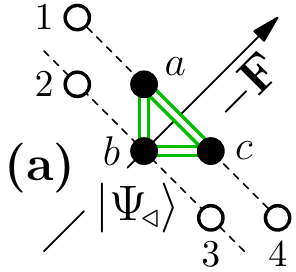}\\[0.25pc]
\hspace*{1pc}\includegraphics[width=0.11\textwidth]{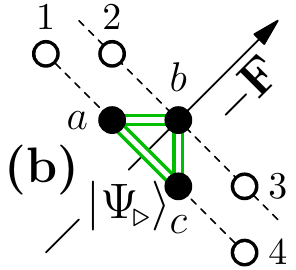}\\[-8.125pc]
\hspace*{6.5pc}\includegraphics[width=0.33\textwidth]{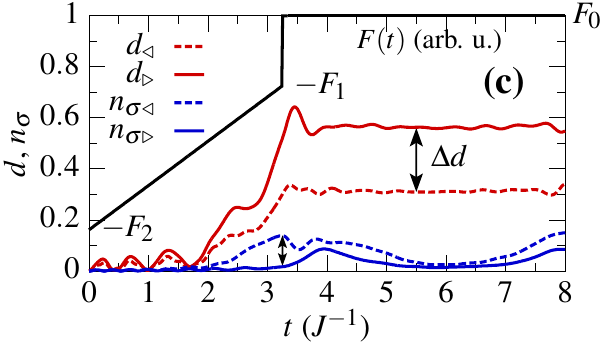}
}
\caption{(a) and~(b): Illustration of the states $\ket{\Psi_{\!\triangleright}}$ and 
$\ket{\Psi_{\!\triangleleft}}$ as defined in the text. (c)~Time evolution of $d_{\triangleright,\triangleleft}(t)$ 
and $n_{\sigma\triangleright,\triangleleft}(t)$ in the negative part of the field cycle $F(t)$ (black 
solid line); the calculations have been performed for an $8\times8$ cluster with the initial states 
$\Psi_{\triangleright,\triangleleft}$ prepared in its center. The field parameters $F_{0,1,2}$ are as 
in \fig{fig:fig4}, $U=10J$ and $\Delta t=3.25 J^{-1}$.
}
\label{fig:fig5}
\end{figure}

To better understand the very different nature of the first and second cycle, we separately 
study the time evolution of states that are present at half the field cycle at times $t=10J^{-1}$ 
and $30J^{-1}$, {\em i.e.}, when the field has just switched sign. Ideally, these states have zero 
double occupancy and form a triangle $abc$ on the lattice which is either facing into the direction $\vec{F}(t)$ 
(triangle $\triangleleft$ in \fig{fig:fig5}a, intermediate state during the first cycle) or 
into the opposite direction (triangle $\triangleright$ in \fig{fig:fig5}b, intermediate state during the second cycle). 
At the avoided crossing $F(t)=-U$, a state with two particles on triangle $\triangleleft$ 
becomes resonantly coupled by two hoppings with states where particles reside {\em outside} the triangle 
(a particle can move from site $b$ to sites $2$ or $3$ via a resonant doublon state on sites $a$ 
and $c$, respectively). A state on triangle $\triangleright$, on the other hand, is only coupled 
resonantly to states {\em within} the triangle, {\em i.e.}, the doublon on site $b$. Hence there is an 
additional channel for coherence loss when state $\ket{\Psi_\triangleleft}$ is moved through 
the crossing, which is always as dominant as the main process of forming the doublon and can 
thus not be avoided. As seen from \fig{fig:fig4}, the coherence loss in the first cycle 
is accompanied by scattering of particles from site $(3,2)$ to site $(3,1)$ and equally 
from $(2,3)$ to $(1,3)$ (see the arrow). To further illustrate the effect, we can prepare wave 
functions $\ket{\Psi_{\!\triangleright,\triangleleft}}=
\frac{1}{\sqrt{6}}\sum_{\cal P} \ket{0,\uparrow,\downarrow}_{\triangleright,\triangleleft}$, where 
$\cal P$ denotes all permutations of the sites $a$, $b$ and $c$, and 
$\mean{n_{\sigma \alpha}}_{\triangleright,\triangleleft}=1/3$ with $\alpha=a,b,c$ (the 
ket-state is encoded as $\ket{a,b,c}$). During the negative part of the field cycle $F(t)$, 
the states $\ket{\Psi_\triangleright}$ and $\ket{\Psi_\triangleleft}$ time evolve as shown 
in \fig{fig:fig5}c. In particular, we find a difference of a factor of two in the total 
double occupancies $d_\triangleright(t)$ and $d_\triangleleft(t)$ for times $t>4J^{-1}$ 
which goes along with an early increase of the density $n_{\sigma\triangleleft}=
\sum_{\beta}\mean{n_{\beta\sigma}}_{\triangleleft}$ at $t\gtrsim2J^{-1}$ in contrast to 
$n_{\sigma\triangleright}=\sum_{\beta}\mean{n_{\beta\sigma}}_{\triangleright}$ ($\beta =1,2,3,4$).

In \fig{fig:fig4}b, we extend the analysis of 
\fig{fig:fig4}a and monitor the transverse charge distribution 
$\mean{n_{i\sigma}^\perp}(\vec{r}_\perp)$  [colored solid lines] over more than two field cycles;
note that the distributions 
are time-averaged over a period of $10J^{-1}$ and have been normalized to $1$. As discussed above, 
the diagonal doublon transport occurs in two stages. In the course of this, $\mean{n_{i\sigma}^\perp}$ 
broadens (during odd cycles) and gets compressed (during even cycles). In particular, we find that 
the degree of compression is such that the transverse decay of the two-particle wave function is smaller 
than the broadening of a binomial distribution $B_{n-1}(\vec{r}_\perp)$, cf.~\fig{fig:fig4}c.
This again elucidates the quantum nature of the ratchet-like motion.

\begin{figure}
\includegraphics[width=0.485\textwidth]{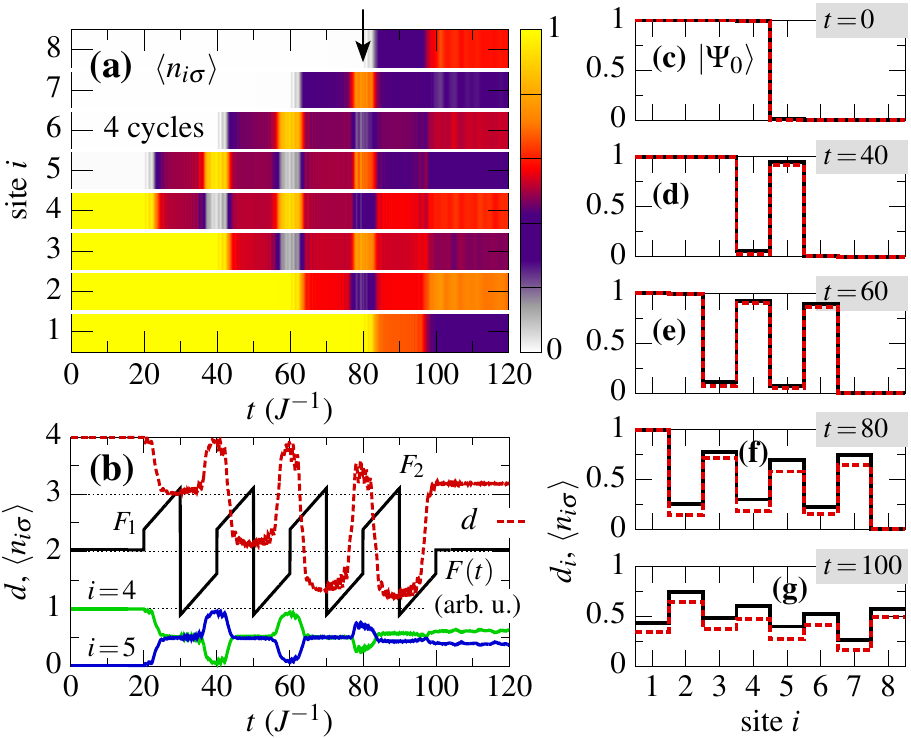}
\caption{Field-assisted manipulation of charge order on an eight-site Hubbard chain 
at $U=20J$ and half-filling; the number of up and down spins is $N_\sigma=\sum_i\mean{n_{i\sigma}}=4$. 
(a)~Time evolution of the densities $\mean{n_{i\sigma}}$ for the system initially (at $t=0$) 
prepared in the band insulating state $\ket{22220000}$, cf.~panel~(c). (b)~Time dependence of double occupation and 
selected densities; the amplitude of the field is indicated by the black solid line. Panels~(c)-(g) 
show the local density $\mean{n_{i\sigma}}$ and double occupation $d_i$ for times $t/J^{-1}=0$, $40$, $60$, $80$ and $100$. 
The field parameters are $F_0=J$, $F_1=12.5J$, $F_1=36J$ and $\Delta t=10J^{-1}$.}
\label{fig:fig6}
\end{figure}

Because the protocol relies on many-particle interactions, it can potentially be used to transform 
more complex states among each other. In the final part of this Letter, we use the protocol 
to transfer a ``band-insulating cluster'' into a charge-ordered configuration. To this end, we start from 
a one-dimensional band insulator modeled by a half-filled eight-site 
Hubbard chain at $U=20J$ which is initially in the Stark eigenstate $\ket{\Psi_0}=\ket{22220000}$ 
for $F_0=J$, {\em i.e.}, the first $4$ sites are doubly occupied and the remaining sites are empty; the initial double occupation is
$d(0)=\sum_i d_i(0)=4$. In figs.~\ref{fig:fig6}a-g, we plot the time evolution of the system 
following four field cycles 
$\vec{F}(t)=F(t)\vec{e}_x$ of the asymmetric form (\ref{eq:field}), with similar field parameters as 
before. As one can see, the first cycle displaces only the doublon which is closest to the 
spatial charge discontinuity whereas all other carriers inside the insulating region (sites $1$ to $3$) 
remain Pauli blocked. Therefore, site $4$ is almost empty at the onset of the second cycle at time 
$t=40J^{-1}$ (\fig{fig:fig6}d), and subsequently two doublons can be moved simultaneously 
by the field, leaving behind two holes on sites $3$ and $5$ at $t=60J^{-1}$ 
(\fig{fig:fig6}e). Consistently the double occupancy drops temporarily by $\Delta d\approx2$ during the 
second cycle. Assisted by the third cycle the initial isolating state is then further transferred into 
one with alternating charge density but still high double occupancy, see the arrow in \fig{fig:fig6}a. 
Aside from some minor loss of coherence the state at $t=80J^{-1}$ is well described 
by the charge-ordered configuration $\ket{20202020}$, compare with \fig{fig:fig6}f. Finally, 
the next cycle shifts the whole charge-ordered configuration by one site to $\ket{02020202}$ (\fig{fig:fig6}g) 
where more than $75$\% of the initial double occupancy, is preserved for the chosen field parameters; 
for optimal control parameters and an extended system (where finite size effects 
become negligible) we expect to observe an even more pronounced signature of this inversion.

In summary, we have studied a quantum ratchet effect in which doublons in the Hubbard model
are moved by unbiased but time-asymmetric electric fields. The effect exploits an adiabatic 
switching through avoided level crossings in the Stark spectrum (for $U/J\gg 1$) and involves 
a complete but temporary fragmentation of the doublon at intermediate times. Though adiabaticity 
is required to avoid dephasing, the mechanism is extremely fast, taking place on time scales 
comparable to the inverse hopping. The process allows for a directional transport of many doublons
simultaneously, and it may be applied to manipulate more complex many-body states. As such, the
protocol is interesting from theoretical perspective because it is associated with fast dynamical
transitions between states of spatially homogeneous and inhomogeneous charge densities. Fields of 
the order $U$ can easily be realized for ultracold atoms in optical lattices~\cite{greiner:02}
such that the proposed manipulation scheme may be a good test-ground for experimental setups 
with single-site resolution. In condensed matter systems, fields corresponding to $F\sim U$ would 
be extremely large, but related protocols might still play a role for the manipulation of complex 
states. For example, the shift of the $\ket{20202020}$ configuration which we have demonstrated 
may be reflected in the rectification of time-asymmetric laser pulses in a charge-ordered medium.

\acknowledgments
We thank \Name{M.~Ganahl} for stimulating discussions. Numerical calculations 
have been performed at the PHYSnet computer cluster at University Hamburg.


\begin{thebibliography}{0}
\bibitem{fn:2}
In order to time evolve the many-body state according to $\ket{\Psi(t)}=U(t,0)\ket{\Psi_0}$ with $U(t',t)=T\exp{(-\ii\int_t^{t'}\mathrm{d}s\,H(s))}$ and time-ordering operator $T$ we use the Krylov method.
\end{thebibliography}
\end{document}